\documentclass[preprintnumbers, floatfix, preprintnumbers, letterpaper, nofootinbib, twocolumn]{revtex4}
\pdfoutput=1
\usepackage{graphicx}
\usepackage{microtype}
\usepackage{amsmath}
\usepackage{amssymb}
\usepackage{subfigure}
\usepackage{hyperref}
\usepackage{url}
\usepackage{xcolor}
\usepackage{color}
\usepackage{mathrsfs}
\usepackage{calrsfs}
\usepackage{amsfonts}
\usepackage{latexsym}
\usepackage{ragged2e}
\usepackage{epsfig}
\usepackage{textcomp}
\usepackage{phaistos}

\makeatletter
\renewcommand\@makefnmark{\hbox{\@textsuperscript{\normalfont\color{purple}\@thefnmark}}}
\renewcommand\@makefntext[1]{%
  \parindent 1em\noindent
            \hb@xt@1.8em{%
                \hss\@textsuperscript{\normalfont\@thefnmark}}#1}
\makeatother

\usepackage{caption}
\DeclareCaptionJustification{justified}{\leftskip=0pt \rightskip=0pt \parfillskip=0pt plus 1fil}
\definecolor{vividviolet}{rgb}{0.62, 0.0, 1.0}
\definecolor{amaranth}{rgb}{0.9, 0.17, 0.31}
\definecolor{palatinateblue}{rgb}{0.15, 0.23, 0.89}
\definecolor{brightpink}{rgb}{1.0, 0.0, 0.5}
\definecolor{cornflowerblue}{rgb}{0.39, 0.58, 0.93}
\definecolor{deepcarminepink}{rgb}{0.94, 0.19, 0.22}
\definecolor{radicalred}{rgb}{1.0, 0.21, 0.37}

\hypersetup{ linktoc=all,
    colorlinks, linkcolor={palatinateblue},
    citecolor={brightpink}, urlcolor={black}
}

\graphicspath{{Images/}}

\graphicspath{{Images/}}



\def\sideremark#1{\ifvmode\leavevmode\fi\vadjust{\vbox to0pt{\vss
 \hbox to 0pt{\hskip\hsize\hskip1em
 \vbox{\hsize1.5cm\tiny\raggedright\pretolerance10000
 \noindent #1\hfill}\hss}\vbox to8pt{\vfil}\vss}}}%
\begin{document}

\title{Momentum-Resolved Probing of Lorentz-Violating Dispersion Relations via Unruh-DeWitt Detector}
\author{Hao Xu}
\thanks{Corresponding author}
\email{haoxu@yzu.edu.cn}
\affiliation{Center for Gravitation and Cosmology, College of Physical Science and Technology, Yangzhou University, \\180 Siwangting Road, Yangzhou City, Jiangsu Province 225002, China}

\begin{abstract}
Inspired by quantum gravity frameworks predicting Planck-scale deviations from Lorentz invariance, we probe Lorentz symmetry violation via modified dispersion relations $\omega_{|\textbf{k}|}$. Departing from conventional approaches, we employ an Unruh-DeWitt detector to probe energy-dependent modifications to the dispersion relations. Two key methodological advances are introduced: (i) a generalized formulation for detector acceleration without assuming specific dispersion relations, and (ii) a momentum-resolved detection paradigm enabling spectral decomposition of $\omega_{|\textbf{k}|}$ through localized momentum-shell integration. Analysis of deviations reveals disruption of the thermal spectrum under significant departures from the Lorentz invariance, while small perturbative regimes manifest as phase-modulated thermal distributions. By restricting detector-field interactions to narrow spectral windows and performing iterative Taylor expansions around reference momenta $|\textbf{k}_0|$, we derive coefficients encoding derivatives of $\omega_{|\textbf{k}|}$, reconstructing its global profile via momentum-space tomography. Our approach offers a scalable method to test Lorentz symmetry violation across energy scales, and establishes a foundation for experimental verification of Planck-scale relics through high-precision spectral measurements.
\end{abstract}

\maketitle

\section{Introduction}

As a cornerstone of Einstein's theory of relativity, Lorentz invariance ensures that physical laws remain invariant across inertial frames. However, several quantum gravity frameworks—such as loop quantum gravity, noncommutative geometry, and Ho\v{r}ava-Lifshitz gravity—suggest that spacetime can acquire discrete or pixelated structures at Planck scale energy $M_p$, potentially violating Lorentz invariance \cite{AmelinoCamelia:2013zea}. The study of Lorentz violation could provide empirical guidance for the unification of quantum mechanics with general relativity, and resolve whether Lorentz invariance is an emergent low-energy phenomenon or an fine-tuning problem \cite{Collins:2004bp,Gambini:2011nx,Polchinski:2011za}. Understanding this could redefine fundamental concepts such as causality, spacetime geometry and the hierarchy of physical scales, while opening up avenues to new theories of quantum spacetime.

A common thread among theories proposing Lorentz violation is their reliance on high-energy departures from Lorentz invariance, coupled with observable signatures at extreme energy scales, while recovering standard symmetry at low energies. In quantum gravity frameworks or modified spacetime models, Lorentz violation typically emerges near the Planck energy, where the fabric of spacetime is hypothesized to become granular, anisotropic, or nonlocal, while at low energies these theories incorporate mechanisms to align with established Lorentz invariant physics. Lorentz invariance may act as an emergent approximation in our observable universe, masking deeper quantum-geometric structures that only become relevant in extreme regimes. Experimental efforts to detect such high-energy relics  aim to bridge the gap between quantum gravity predictions and empirical reality.

While the Planck energy scale $ (\sim 10^{19}\ \mathrm{GeV}) $ vastly exceeds currently attainable experimental energies, modifications to Lorentz-invariant dispersion relations often employ phenomenological extensions through higher-order momentum terms. A conventional approach introduces successive momentum corrections to the standard energy-momentum relation:
\begin{align}
\omega_{|\textbf{k}|}^2 = |\textbf{k}|^2 + m^2 + \frac{\kappa_1}{M_p}|\textbf{k}|^3 + \frac{\kappa_2}{M_p^2}|\textbf{k}|^4 + \mathcal{O}(|\textbf{k}|^5),
\label{eq:dispersion}
\end{align}
where $\{\kappa_i\}$ are dimensionless coefficients parametrizing Lorentz violation. The modified dispersion relation produces kinematic constraints on energy-momentum conservation that are different from the usual Lorentz-invariant case, so that reactions that are normally forbidden may occur, and thresholds for reactions are modified. 

\begin{figure}
\begin{center}
\includegraphics[width=0.45\textwidth]{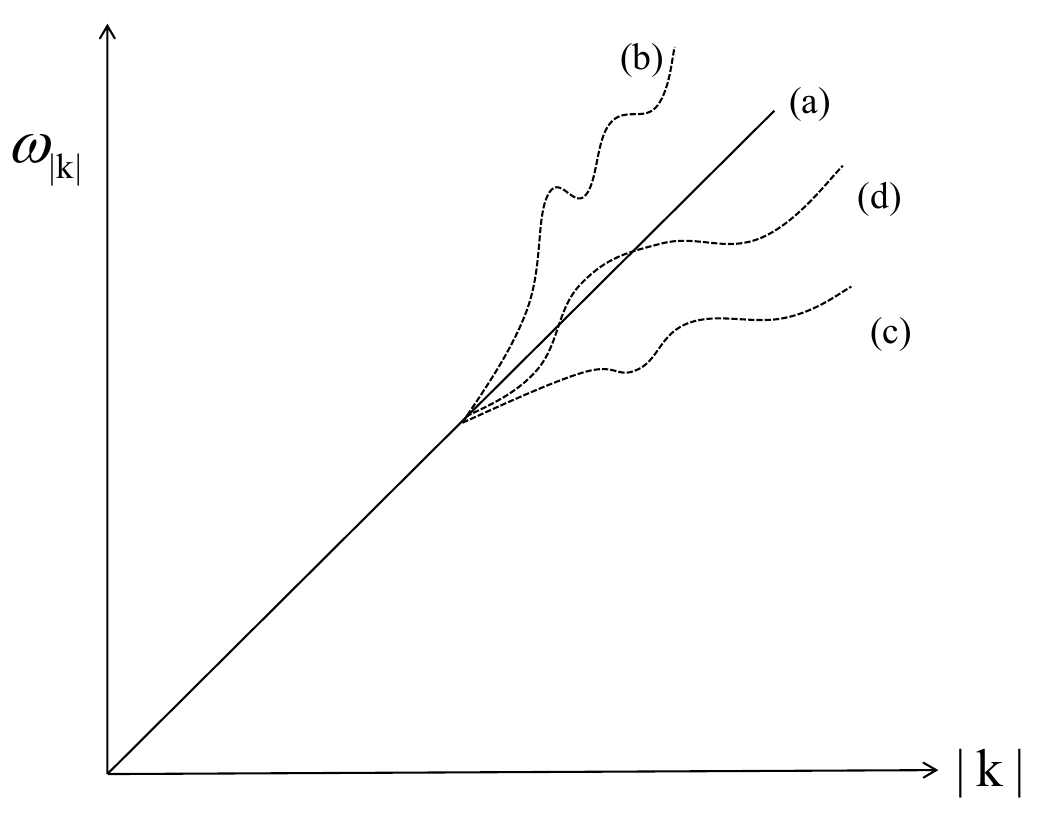}
\caption{Schematic diagram of the dispersion relation for free massless scalar field. The four curves in the diagram correspond to the four cases of $\omega_{|\textbf{k}|}$: (a). The Lorentz invariance case $\omega_{|\textbf{k}|}= |\textbf{k}|$. (b). The $\omega_{|\textbf{k}|}$ is always larger than $|\textbf{k}|$. (c). The $\omega_{|\textbf{k}|}$ is always smaller than $|\textbf{k}|$. (d). At different $|\textbf{k}|$, there are $\omega_{|\textbf{k}|}> |\textbf{k}|$ or $\omega_{|\textbf{k}|}< |\textbf{k}|$. }
\label{fig1}
\end{center}
\end{figure}

However, in the absence of a specific theoretical framework, Lorentz violations in the dispersion relation could manifest through diverse relationships between $\omega_{|\textbf{k}|}$ and $|\textbf{k}|$. Physical theories often exhibit distinct behaviors across energy scales, leaving the functional form of $\omega_{|\textbf{k}|}$ unresolved. For a free massless scalar field, it is well-established that $\omega_{|\textbf{k}|} \approx |\textbf{k}|$ holds at sufficiently small $|\textbf{k}|$ values. As $|\textbf{k}|$ increases, this proportionality becomes uncertain. The function $\omega_{|\textbf{k}|}$ might preserve Lorentz invariance by maintaining strict equality with $|\textbf{k}|$. Alternatively, it could systematically exceed or fall below $|\textbf{k}|$ beyond a critical threshold, or even alternate between regimes where $\omega_{|\textbf{k}|} > |\textbf{k}|$ and $\omega_{|\textbf{k}|} < |\textbf{k}|$ across different $|\textbf{k}|$ ranges. Fig.\ref{fig1} schematically illustrates these potential configurations under theoretical uncertainties.

One of the central goals of Lorentz violation research is to establish the precise dispersion relation $\omega_{|\textbf{k}|}$ through rigorous theoretical modelling and experimental verification. This pursuit drives multidisciplinary investigations ranging from photon decay/annihilation, vacuum \v{C}erenkov effect, gravitational coupling and additional wave modes in high energy physics \cite{Jacobson:2002hd,Jacobson:2005bg} to condensed matter analyses of quantum phase transitions in dipolar Bose-Einstein condensates \cite{Tian:2021mka,Tian:2022gfa}. Our work departs from conventional approaches by implementing an Unruh-DeWitt detector within the quantum field-theoretic framework of Lorentz violation, thereby deriving novel predictions for the $\omega_{|\textbf{k}|}$ functional form. As an operational tool bridging quantum field theory and gravitational physics, the Unruh-DeWitt detector allows a unified analysis of Lorentz violation effects across energy scales, and its sensitivity to modified vacuum fluctuations provides critical insights into how Lorentz violation reshapes our fundamental understanding of spacetime geometry and gravitational interactions.

While existing studies have addressed aspects of Unruh-DeWitt detector dynamics in Lorentz-violating frameworks, this work introduces two methodological advances: (i) We establish a generalised formulation for detector acceleration without assuming specific forms of the dispersion relation $\omega_{|\textbf{k}|}$, in contrast to previous approaches that either impose fixed $\omega_{|\textbf{k}|}$ parameterisations \cite{Davies:2023zfq,Rinaldi:2007de,Rinaldi:2008qt} or restrict the analysis to uniform/circular motion \cite{Husain:2015tna,Louko:2017emx,JaffinoStargen:2017qyh}. (ii) By developing a momentum-resolved detection paradigm rather than conventional time-dependent transition rates, we enable mode-specific feedback analysis through localised spectral decomposition. This involves expanding $\omega_{|\textbf{k}|}$ around reference momenta $|\textbf{k}_0|$, integrating momentum shell responses to extract dispersion coefficients, and iteratively reconstructing the global $\omega_{|\textbf{k}|}$ profile via stepwise ${|\textbf{k}_0|}$ scans, a strategy analogous to momentum-space tomography that resolves anisotropies inaccessible to temporal averaging techniques.

\section{The model of Unruh-DeWitt detector}

The Unruh-DeWitt detector model consists of a two-level quantum system (detector) interacting with a quantized field. Originally proposed to study the Unruh effect \cite{Unruh:1976db,Unruh:1983ms,DeWitt1979}, which predicts that a uniformly accelerated observer perceives the Minkowski vacuum as a thermal state, the system is governed by the interaction Hamiltonian
\begin{equation}
\hat{H}_{\text{int}} = \lambda \hat{\sigma}_x(\tau) \hat{\phi}[x(\tau)] \propto (\hat{\sigma}_{-} + \hat{\sigma}_{+})(\hat{a} + \hat{a}^{\dagger}),
\end{equation}
where $\lambda$ denotes the coupling constant, $\hat{\sigma}_{+} = |e\rangle\langle g|$ and $\hat{\sigma}_{-} = |g\rangle\langle e|$ are the detector's excitation and de-excitation operators, while $\hat{a}$ and $\hat{a}^{\dagger}$ represent the field's annihilation and creation operators respectively. This model provides a simplified description of light-matter interaction phenomena. The interaction Hamiltonian contains four distinct terms corresponding to different physical processes: 
\begin{itemize}
\item The rotating wave terms $\hat{\sigma}_{+}\hat{a} + \hat{\sigma}_{-} \hat{a}^{\dagger}$ describe stimulated emission/absorption processes, reducing to the Jaynes-Cummings model under the rotating wave approximation \cite{Jaynes1963}.
\item The counter-rotating wave terms $\hat{\sigma}_{+} \hat{a}^{\dagger} + \hat{\sigma}_{-}\hat{a}$ are responsible for the Unruh effect and its counterpart.
\end{itemize}

Notably, when the detector resides in its ground state and the field remains in the vacuum state, only the specific term $\hat{\sigma}_{+} \hat{a}^{\dagger}$ associated with the Unruh effect contributes to the system's dynamics. In the interaction picture, the operator evolves as
\begin{equation}
\hat{\sigma}_x(\tau)=\hat{\sigma}^{+}e^{i\Omega \tau}+\hat{\sigma}^{-}e^{-i\Omega \tau},
\end{equation}
and the field operator admits the mode expansion
\begin{align}
\hat{\phi}[x(\tau)] = \int \tilde{\text{d}} \textbf{k} &\left[ \hat{a}(\textbf{k})e^{-i\omega_{|\textbf{k}|} t(\tau)+i{\textbf{k}\cdot{\textbf{x}}(\tau)}} \right. \nonumber \\
&\left. +\hat{a}^{\dagger}(\textbf{k})e^{i\omega_{|\textbf{k}|} t(\tau)-i{\textbf{k}\cdot{\textbf{x}}(\tau)}} \right],
\label{int}
\end{align}
where the invariant measure $\tilde{\text{d}}\textbf{k}\equiv\frac{\text{d}^n \textbf{k}}{\sqrt{(2\pi)^n 2\omega_{|\textbf{k}|}}}$. The $\hat{\sigma}_{+} \hat{a}^{\dagger}$ term embodies simultaneous excitation processes: the detector transitions to its excited state while generating a field quantum. 

To quantify the detector's response, we construct the reduced density matrix of the detector by tracing out the field degrees of freedom. The excited state population emerges as
\begin{align}
\int \frac{\text{d}^n \textbf{k}}{{(2\pi)^n}2\omega_{|\textbf{k}|}}\left|\int \text{d}\tau\, e^{i\Omega \tau}e^{i\omega_{|\textbf{k}|} t(\tau)-i{\textbf{k}\cdot{\textbf{x}}(\tau)}}\right|^2,
\end{align}
which encodes both the detector's internal dynamics $\Omega \tau$ and the field mode propagation $\omega_{|\textbf{k}|} t(\tau)-\textbf{k}\cdot{\textbf{x}}(\tau)$.

However, when integrating over the full momentum space $\textbf{k} \in \mathbb{R}^n$ and the detector's proper time $\tau \in \mathbb{R}$, the expression necessarily diverges. This divergence arises fundamentally because coupling to all momentum modes enables instantaneous excitations within finite proper time intervals, and the double temporal integration $\iint d\tau d\tau'$ in the transition amplitude inevitably accumulates infinite phase contributions. The standard regularization procedure involves sequential integration \cite{Davies}: first compute the momentum-space integral to obtain the field's two-point correlation function $\langle \hat{\phi}(\tau)\hat{\phi}(\tau') \rangle$, then evaluate the relative-time integral over $\Delta \tau \equiv \tau - \tau'$ to derive the time-averaged transition rate per $ (\tau + \tau')/2$.

The mutual excitation of detector and field constitutes a quantum correlated process with small probability amplitude, and its experimental observation is direct verification of the Unruh effect. The existing research on the experimental aspects of the Unruh effect has focused on amplification of the effect, which is challenging. To clarify these constraints, consider a $(1+1)$-dimensional Lorentz-invariant free massless scalar field coupled to an inertial detector. The detector's spacetime trajectory $x^\mu(\tau) = (\gamma \tau, \gamma \textbf{v}\tau)$ with $\gamma=(1-|{\textbf{v}}|^2)^{-1/2}$ induces the phase integral:
\begin{align}
\int \text{d}\tau e^{i\Omega \tau}e^{i\omega_{|\textbf{k}|} t(\tau)-i{\textbf{k}\cdot{\textbf{x}}(\tau)}}\propto \delta\left(\Omega + \gamma(\omega_{|\textbf{k}|} - \textbf{k}\cdot \textbf{v}) \right).
\end{align}
The Lorentz-invariant dispersion relation $\omega_{|\textbf{k}|} = |\textbf{k}|$ guarantees $\omega_{|\textbf{k}|} - \textbf{k}\cdot \textbf{v} \geq 0$, forcing the $\delta$-function to vanish identically. This null result reflects the energy-momentum conservation constraints in Poincaré-symmetric systems.

Remarkably, Lorentz symmetry violation modifies this paradigm. If modified dispersion relations permit subluminal phase velocities ($\omega_{|\textbf{k}|}/|\textbf{k}| < 1$) within certain momentum regimes, the critical condition $\gamma(\omega_{|\textbf{k}|} - \textbf{k} \cdot \textbf{v}) = -\Omega$ becomes achievable \cite{Husain:2015tna,Louko:2017emx}. When the detector's rapidity exceeds thresholds, spontaneous excitations emerge as measurable resonance peaks.

However, when the dispersion relation satisfies $\omega_{|\textbf{k}|}/|\textbf{k}| > 1$, the inertial detector exhibits vanishing response. This necessitates analysis of a uniformly accelerated detector. Focusing on $1+1$ dimensions, which aligns with the prediction in many quantum gravity theories that spacetime dimensionality is dynamical and scale-dependent, flowing toward two dimensions at high energy scales \cite{Carlip:2017eud}, we parametrize the detector's Rindler trajectory as $x^\mu(\tau) = \left( \frac{1}{a}\sinh(a\tau),\ \frac{1}{a}\cosh(a\tau) \right)$, where $a$ denotes proper acceleration. The excitation amplitude decomposes into positive/negative frequency components:
\begin{align}
f_{\pm}(|\textbf{k}|)=\int^{\infty}_{-\infty} \text{d}\tau\, e^{i\Omega \tau}e^{i\omega_{|\textbf{k}|} t(\tau)\pm i|\textbf{k}| x(\tau)}.
\label{fk}
\end{align}
Substituting the trajectory components, the $\tau$-integration yields:
\begin{align}
f_{\pm}(|\textbf{k}|)=\frac{2}{a}e^{-\frac{\pi \Omega}{2a}}\left( \frac{\omega_{|\textbf{k}|} - |\textbf{k}|}{\omega_{|\textbf{k}|} + |\textbf{k}|} \right)^{\pm i\frac{\Omega}{2a}}K_{i\Omega/a}\!\left(\frac{\sqrt{\omega_{|\textbf{k}|}^2 - |\textbf{k}|^2}}{a} \right),
\label{fk2}
\end{align}
where $K_{\nu}(z)$ denotes the modified Bessel function of the second kind. 

Under Lorentz-invariant dispersion relations ($\omega_{|\textbf{k}|} = |\textbf{k}|$), the excitation amplitude simplifies to
\begin{align}
f_{\pm}(|\textbf{k}|) = \frac{1}{a}\left(\frac{a}{|\textbf{k}|}\right)^{\pm i\frac{\Omega}{a}} e^{-\frac{\pi \Omega}{2a}} \Gamma\!\left(\pm i\frac{\Omega}{a}\right),
\label{fk3}
\end{align}
where $\Gamma(z)$ denotes the Euler gamma function. Utilizing the modulus identity for gamma function
\begin{equation}
|\Gamma(\pm ix)|^2 = \frac{\pi}{x \sinh(\pi x)},
\end{equation}
we have
\begin{align}
|f_{\pm}(|\textbf{k}|)|^2 = \frac{2\pi}{a\Omega} \frac{1}{e^{2\pi \Omega/a} - 1}.
\end{align}
Notably, the absence of $|\textbf{k}|$-dependence reflects thermalization of all field modes, and the Bose-Einstein distribution factor $(e^{2\pi \Omega/a}-1)^{-1}$ confirms thermal response at Unruh temperature $T_U = a/(2\pi)$.

The thermality of the Unruh effect crucially depends on the linear dispersion relation $\omega_{|\textbf{k}|} = |\textbf{k}|$. Significant deviations from this relation would destroy the thermal character of the detector's response, whereas perturbatively small modifications may preserve thermality within the regime of validity of perturbation theory. Crucially, the term $(\omega_{|\mathbf{k}|} - |\mathbf{k}|)$ governs the asymptotic behavior of the integrand of \eqref{fk}, revealing a fundamental mathematical distinction between the dispersion relations $\omega_{|\mathbf{k}|} = |\mathbf{k}|$ and $\omega_{|\mathbf{k}|} \neq |\mathbf{k}|$. Notably, even if we consider the limit case $\omega_{|\mathbf{k}|} \to |\mathbf{k}|$, a rapidly oscillating phase term must be introduced. A rigorous analysis of this behavior is provided in the appendix.

A critical challenge is that integration over the full momentum space inherently averages out scale-dependent features of $\omega_{|\textbf{k}|}$, yielding a coarse-grained description. To probe the energy-scale dependence of $\omega_{|\textbf{k}|}$, one must implement a spectral decomposition approach by restricting integration to narrow momentum shells. This corresponds to coupling the detector selectively to field modes within a spectral window, thereby enabling momentum-resolved measurement of dispersion relations.

Defining 
\begin{align}
g(|\textbf{k}|)&= \frac{1}{{(2\pi)}2\omega_{|\textbf{k}|}}|f_{\pm}({|\textbf{k}|})|^2 \nonumber \\
&=\frac{1}{\pi \omega_{|\textbf{k}|} a^2}e^{-\frac{\pi \Omega}{a}}\left| K_{i\Omega/a}\left(\frac{\sqrt{\omega_{|\textbf{k}|}^2-|\textbf{k}|^2}}{a} \right) \right|^2,
\label{gk}
\end{align}
we can observe the amplitude symmetry between positive and negative frequency components ($\pm$ sign selection) reveals a fundamental degeneracy in the detector's response. This is quantified through the invariant transition probability $|f_{+}(|\textbf{k}|)|^2 = |f_{-}(|\textbf{k}|)|^2$. It eliminates redundant degrees of freedom associated with directional coordinates, focusing analysis on energy-scale dependence, and enables universal treatment of dispersion relations through scalar parameterization. 

\section{Spectral decomposition via localized momentum windows}

To analyze the detector's spectral response near a reference momentum $|\textbf{k}_0|$, we perform Taylor expansions of key functions. The dispersion relation $\omega_{|\textbf{k}|}$ expands to second order as
\begin{align}
\omega_{|\textbf{k}|} \approx \omega_0 + \omega_1 \Delta \textbf{k} + \frac{\omega_2}{2}(\Delta \textbf{k})^2 + \mathcal{O}\!\left( (\Delta \textbf{k})^3 \right),
\end{align}
where $\omega_0 \equiv \omega_{|\textbf{k}_0|}$, $\Delta \textbf{k} \equiv |\textbf{k}| - |\textbf{k}_0|$, with $\omega_1$ and $\omega_2$ denoting first and second derivatives of $\omega_{|\textbf{k}|}$ at $|\textbf{k}|=|\textbf{k}_0|$. The expansion for $1/\omega_{|\textbf{k}|}$ becomes
\begin{align}
\frac{1}{\omega_{|\textbf{k}|}} \approx \frac{1}{\omega_0} - \frac{\omega_1}{\omega_0^2}\Delta \textbf{k} + \left( \frac{\omega_1^2}{\omega_0^3} - \frac{\omega_2}{2\omega_0^2} \right)(\Delta \textbf{k})^2 + \mathcal{O}\!\left( (\Delta \textbf{k})^3 \right).
\end{align}

For the parameter $z(|\textbf{k}|) \equiv \sqrt{\omega_{|\textbf{k}|}^2 - |\textbf{k}|^2}/a$, the second-order expansion yields
\begin{align}
z(|\textbf{k}|) \approx z_0 + z_1\Delta \textbf{k} + z_2(\Delta \textbf{k})^2,
\end{align}
with coefficients determined by
\begin{align}
z_0 &= \frac{\sqrt{\omega_0^2 - |\textbf{k}_0|^2}}{a}, \\
z_1 &= \frac{\omega_0\omega_1 - |\textbf{k}_0|}{a\sqrt{\omega_0^2 - |\textbf{k}_0|^2}}, \\
z_2 &= \frac{(\omega_1^2 + \omega_0\omega_2 - 1)(\omega_0^2 - |\textbf{k}_0|^2) - (\omega_0\omega_1 - |\textbf{k}_0|)^2}{2a(\omega_0^2 - |\textbf{k}_0|^2)^{3/2}}.
\end{align}

The modified Bessel function $K_{i\Omega/a}(z(|\textbf{k}|))$ subsequently expands as
\begin{align}
K_{i\Omega/a}(z(|\textbf{k}|)) &\approx K_{i\Omega/a}(z_0) + \left[z_1 K'_{i\Omega/a}(z_0)\right]\Delta \textbf{k} \nonumber \\
&\quad + \left[z_2 K'_{i\Omega/a}(z_0) + \frac{z_1^2}{2}K''_{i\Omega/a}(z_0)\right](\Delta \textbf{k})^2,
\end{align}
where the first and second order derivatives of $K_{\nu}(z_0)$ satisfy
\begin{align}
K'_{\nu}(z_0) &= -K_{\nu+1}(z_0) + \frac{\nu}{z_0} K_{\nu}(z_0), \nonumber \\
K''_{\nu}(z_0) &= \left( 1 + \frac{\nu^2}{z_0^2} \right) K_{\nu}(z_0) - \frac{1}{z_0} K'_{\nu}(z_0). \nonumber
\end{align}

The spectral coupling function $g(|\textbf{k}|)$ then admits the quadratic approximation
\begin{equation}
g(|\textbf{k}|) \approx \mathcal{C}_0 + \mathcal{C}_1\Delta \textbf{k} + \mathcal{C}_2(\Delta \textbf{k})^2,
\end{equation}
with coefficients
\begin{align*}
\mathcal{C}_0 &= \frac{e^{-\pi\Omega/a}}{\pi a^2\omega_0}K_0^2, \\
\mathcal{C}_1 &= \frac{e^{-\pi\Omega/a}}{\pi a^2}\left(-\frac{\omega_1}{\omega_0^2}K_0^2 + \frac{2z_1}{\omega_0}K_0 K'_0\right), \\
\mathcal{C}_2 &= \frac{e^{-\pi\Omega/a}}{\pi a^2}\left[\frac{\omega_1^2}{\omega_0^3}K_0^2 - \frac{\omega_2}{\omega_0^2}K_0^2 - \frac{2\omega_1z_1}{\omega_0^2}K_0 K'_0 \right. \\
&\quad \left. + \frac{1}{\omega_0}\left(2z_2K_0K'_0 + z_1^2\left(K_0 K''_0 + (K'_0)^2\right)\right)\right],
\end{align*}
where $K_0 \equiv K_{i\Omega/a}(z_0)$, $K'_0 \equiv K'_{i\Omega/a}(z_0)$, $K''_0 \equiv K''_{i\Omega/a}(z_0)$, and we utilize the fact that $K_{\nu}(z)$ is a real-valued function when $\nu$ is purely imaginary and $z$ is real.

When restricting detector-field interaction to a finite spectral window $|\textbf{k}| \in [|\textbf{k}_0| - m, |\textbf{k}_0| + n]$, the detector's response becomes the integrated spectral density:
\begin{align}
\int^{|\textbf{k}_0|+n}_{|\textbf{k}_0|-m} g(|\textbf{k}|)\, \text{d} |\textbf{k}| &= \mathcal{C}_0(m + n) + \frac{\mathcal{C}_1}{2}(n^2 - m^2) \nonumber \\
&+ \frac{\mathcal{C}_2}{3}(n^3 + m^3).
\end{align}
This structured response contains three experimentally distinguishable components: 
\begin{itemize}
\item The zeroth-order term $\propto \mathcal{C}_0$ quantifies baseline thermal noise.
\item The linear gradient term $\propto \mathcal{C}_1$ encodes spectral tilt.
\item The curvature term $\propto \mathcal{C}_2$ captures local nonlinearities.
\end{itemize}

By systematically varying the asymmetric window parameters $(m, n)$ under the constraints:
\begin{equation}
m \neq n,\quad \max(m,n) \ll |\textbf{k}_0|,
\end{equation}
one can solve the system for $\{\mathcal{C}_0, \mathcal{C}_1, \mathcal{C}_2\}$, and these coefficients determine the dispersion relation derivatives $\{\omega_0, \omega_1, \omega_2\}$ at $|\textbf{k}_0|$. Sequential application of this protocol across adjacent $|\textbf{k}_0|$ values enables piecewise reconstruction of $\omega_{|\textbf{k}|}$ via concatenated Taylor approximations. The complete functional form of $\omega_{|\textbf{k}|}$ emerges through this incremental spectral deconvolution procedure.

\section{Discussion}

The proposed methodology for reconstructing Lorentz-violating dispersion relations through momentum-resolved Unruh-DeWitt detectors introduces conceptual and technical challenges requiring innovative solutions. A primary obstacle lies in engineering detector-field interactions localized to narrow momentum shells. While standard detectors couple universally to all modes through spatial smearing, achieving spectral selectivity demands non-trivial modifications such as comb-based radio-frequency photonic filters \cite{Supradeepa2012,pasquazi2018micro}. Furthermore, since the Unruh effect has not yet been observed experimentally, one of the most important issues is to enhance the contribution of the observed effect, such as compressing the rotational wave term by changing the kinematics of the detector and to increase the value of the counter-rotational wave term by increasing the number of particles in the quantum field \cite{Kempf2022}. Moreover, the integration over $\tau$ in this work is performed over the entire domain. In practical implementations, we naturally need to introduce a window function and account for its corrections. The discussion on the Unruh effect in \cite{Sriramkumar1996} can be applied to our calculations. In this work, we consider linear acceleration motion in $1+1$-dimensional spacetime, we can also move to $2+1$ or higher dimensions to consider circular motion \cite{Akhmedov:2023qmg}. An interesting approach is to study circular motion in spacetime with black holes \cite{Hodgkinson:2014iua}, where the response of the Unruh–DeWitt detector depends not only on the motion trajectory, but also on the vacuum it is coupled to, which could form the basis of potential extensions.

Future investigations should prioritize experimental feasibility studies using tabletop quantum simulators \cite{Carney:2018ofe}. The framework naturally extends to multi-detector networks for mapping spatial correlations of Lorentz-violating vacuum, and to curved spacetime geometries where dispersion anomalies interact with gravitational redshifts. Long-term prospects involve developing iterative momentum-shell tomography to reconstruct $\omega_{|\textbf{k}|}$ up to Planck-scale, potentially revealing connections between modified dispersion relations and black hole information paradox. Ultimately, the synthesis of quantum detector models with momentum-space resolution opens new avenues for spectral analysis of quantum spacetime \cite{Hossenfelder:2012jw}, and can help us understand the problem of exchanging energy and information when the observer and the spacetime background act as open systems \cite{Xu:2021buk,Xu:2024ztq}. Realizing this vision demands advances in relativistic quantum control, and cross-disciplinary integration of quantum gravity phenomenology with quantum information techniques. The path forward lies in transforming theoretical constructs into laboratory-operational probes through coordinated developments in detector engineering, quantum measurement theory, and relativistic analog simulation platforms.

\begin{acknowledgments}
Hao Xu thanks National Natural Science Foundation of China (No.12205250) for funding support.
\end{acknowledgments}

\section*{Appendix: The Limit Case $\omega_{|\mathbf{k}|} \to |\mathbf{k}|$}
We naturally expected that under large deviations, the thermal spectrum is disrupted, and in the limiting case where $\omega_{|\mathbf{k}|} \rightarrow |\mathbf{k}|$, \eqref{fk2} would degenerate into \eqref{fk3}, thereby preserving thermality under small perturbations. Although discrepancy between \eqref{fk2} and \eqref{fk3} confirms the former's validity,  \emph{the limiting case of $\omega_{|\mathbf{k}|} \to |\mathbf{k}|$ exhibits greater complexity than anticipated}. In this appendix, we analyze the limit case $\omega_{|\mathbf{k}|} \to |\mathbf{k}|$. Although one might expect formula \eqref{fk2} to reduce directly to formula \eqref{fk3} in this limit, this is not mathematically justified. 

The distinction arises from the trajectory $x^\mu(\tau) = \left( \frac{1}{a}\sinh(a\tau),\ \frac{1}{a}\cosh(a\tau) \right)$ in the integral of (7). After the substitution $x = e^{a\tau}$, the expression becomes:
\begin{align*}
\frac{1}{a} \int_{0}^{+\infty} x^{\pm i \Omega/a - 1} 
\exp\left[ \pm i \left( 
\frac{\omega_{|\mathbf{k}|} + |\mathbf{k}|}{2a} x - 
\frac{\omega_{|\mathbf{k}|} - |\mathbf{k}|}{2a x}
\right) \right] \text{d}x.
\end{align*}
Crucially, the term $(\omega_{|\mathbf{k}|} - |\mathbf{k}|)$ governs the asymptotic behavior $x\rightarrow 0^+$ of the integrand. \emph{This demonstrates the fundamental mathematical distinction between the cases $\omega_{|\mathbf{k}|} = |\mathbf{k}|$ and $\omega_{|\mathbf{k}|} \neq |\mathbf{k}|$}, as further evidenced by the oscillatory factor $\left( \frac{\omega_{|\mathbf{k}|} - |\mathbf{k}|}{\omega_{|\mathbf{k}|} + |\mathbf{k}|} \right)^{\pm i\Omega/(2a)}$ in \eqref{fk2}.

To elucidate this behavior, we expand equation \eqref{fk2} using properties of special functions. Parallel analyses are also discussed in \cite{Hammad:2021rey}. Our derivation relies on Bessel function identities from \cite{Gradshteyn}, with specific equation numbers provided for verification.

The modified Bessel function $K_\nu(z)$ relates to Hankel functions via \cite[Eq.~8.407.1]{Gradshteyn}:
\begin{align*}
K_{\nu}(z) = \frac{\pi i}{2}e^{\pi i \nu/2} H_{\nu}^{(1)}\left(ze^{i\pi/2}\right).
\end{align*}
The Hankel function admits the decomposition \cite[Eq.~8.405.1]{Gradshteyn}:
\begin{align*}
H_{\nu}^{(1)}(z) = J_{\nu}(z) + i Y_{\nu}(z),
\end{align*}
where $J_\nu(z)$ and $Y_\nu(z)$ denote Bessel functions of the first and second kind, respectively. These satisfy \cite[Eq.~8.403.1]{Gradshteyn}:
\begin{align*}
Y_{\nu}(z) = \frac{\csc(\nu\pi)}{\pi} \left[ \cos(\nu\pi) J_{\nu}(z) - J_{-\nu}(z) \right].
\end{align*}
The series expansion for $J_\nu(z)$ is given by \cite[Eq.~8.402]{Gradshteyn}:
\begin{align*}
J_{\nu}(z) = \left(\frac{z}{2}\right)^{\nu} \sum_{k=0}^{\infty} \frac{(-1)^k}{k! \, \Gamma(\nu + k + 1)} \left(\frac{z}{2}\right)^{2k}.
\end{align*}
Retaining only the $k=0$ dominant term for small $z$, we obtain:
\begin{align*}
H_{\nu}^{(1)}(iz) &= \frac{e^{\pi i \nu/2} z^{\nu}}{2^{\nu}} \left[ \frac{1 + i\cot(\pi\nu)}{\Gamma(1+\nu)} - \frac{i (2/(iz))^{2\nu} \csc(\pi\nu)}{\Gamma(1-\nu)} \right] \\
&= -\frac{i e^{-\pi i \nu/2}}{\pi} \left[ \Gamma(-\nu) \left(\frac{z}{2}\right)^{\nu} + \Gamma(\nu) \left(\frac{z}{2}\right)^{-\nu} \right].
\end{align*}
Substituting these results into $K_\nu(z)$ and equation \eqref{fk2} yields:
\begin{align*}
f_{\pm}(|\mathbf{k}|) &= \frac{e^{-\pi \Omega/(2a)}}{a} \Bigg[ 
\Gamma\left( \mp i \frac{\Omega}{a} \right) \left( \frac{\omega_{|\mathbf{k}|} - |\mathbf{k}|}{2a} \right)^{\pm i \Omega/a} \\
&\quad + \Gamma\left( \pm i \frac{\Omega}{a} \right) \left( \frac{\omega_{|\mathbf{k}|} + |\mathbf{k}|}{2a} \right)^{\mp i \Omega/a} \Bigg].
\end{align*}
When $\omega_{|\mathbf{k}|} = |\mathbf{k}|$, the first term vanishes and the expression reduces to equation \eqref{fk3}. However, as $\omega_{|\mathbf{k}|} \to |\mathbf{k}|$ from arbitrary values, the first term exhibits singular behavior characterized by rapid phase oscillations. Consequently, for cases where $\omega_{|\mathbf{k}|} \neq |\mathbf{k}|$, the resulting spectrum manifests as a phase-modulated thermal distribution \cite{Hammad:2021rey}:
\begin{align*}
|f_{\pm}(\Omega)|^{2} \approx \frac{8\pi}{a\Omega \left( e^{2\pi\Omega/a} - 1 \right)} \cos^{2}\left[ \theta - \frac{\Omega}{2a} \ln\left( \frac{\omega_{|\mathbf{k}|}^2 - |\mathbf{k}|^2}{4a^2} \right) \right],
\end{align*}
where $\theta = \arg \Gamma(\pm i \frac{\Omega}{a})$. The large $z$ limit of $K_\nu(z)$ were also analysed in \cite{Hammad:2021rey}, and as we expected, the thermal signature is suppressed.



\end{document}